\documentclass[USenglish,online,preprint]{article}

\usepackage[utf8]{inputenc}
\usepackage{microtype}
\usepackage{authblk}

\usepackage[numbers]{natbib}
\usepackage{siunitx} 
\usepackage{graphicx}
\usepackage{dcolumn}
\usepackage{gensymb}
\usepackage{makecell,tabularx}
\usepackage{physics}
\usepackage{appendix}

\usepackage{xr-hyper}
\usepackage{hyperref}
\usepackage{cleveref}

\makeatother

\begin{document}

\title{Discovery of ST1 centers in natural diamond}


\author[2,3,*]{Priyadharshini ~Balasubramanian$^\dagger$}
\author[2,3,*]{Mathias~H.~Metsch$^\dagger$}
\author[1,*]{Prithvi~Reddy$^\dagger$}
\author[4,5]{Lachlan~J.~Rogers}
\author[1]{Neil~B.~Manson}
\author[1]{Marcus~W.~Doherty}
\author[2,3]{Fedor~Jelezko}

\affil[*] {
    These authors contributed equally to this work.
}

\affil[1]{%
    Laser Physics Centre, Research School of Physics and Engineering, Australian National University, Acton, 2601, Australia
}
\affil[2]{%
    Institute for Quantum Optics, University Ulm, Albert-Einstein-Allee 11, D-89081 Ulm, Germany
}%
\affil[3]{%
    Center for Integrated Quantum Science and Technology (IQST), University Ulm, Albert-Einstein-Allee 11, D-89081 Ulm, Germany
}%
\affil[4]{
    Department of Physics and Astronomy, Macquarie University, New South Wales 2109, Australia
}
\affil[5]{
    ARC Centre of Excellence for Engineered Quantum Systems (EQUS)
}


\maketitle

\externaldocument{./supp}

\abstract{
    The ST1 center is a point defect in diamond with bright fluorescence and a mechanism for optical spin initialization and readout. The center has impressive potential for applications in diamond quantum computing as a quantum bus to a register of nuclear spins. This is because it has an exceptionally high readout contrast and, unlike the well-known nitrogen-vacancy center, it does not have a ground state electronic spin that decoheres the nuclear spins. 
    However, its chemical structure is unknown and there are large gaps in our understanding of its properties. We present the discovery of ST1 centers in natural diamond. Our experiments identify interesting power dependence of the center's optical dynamics and reveal new electronic structure. We also present a theory of its electron-phonon interactions, which we combine with previous experiments, to shortlist likely candidates for its chemical structure.
}

\section{Introduction}
When designing hybrid quantum technologies, we seek to develop an architecture that uses different types of qubits \cite{kurizki2015quantum}. The advantage of this approach is that we can exploit each system's strengths while mitigating their weaknesses. By using a hybrid architecture, the diamond quantum computing platform demonstrates impressive performance at room temperature \cite{Waldherr2014}. In particular, this application uses negatively-charged nitrogen-vacancy (NV$^-$) center's electron spin as a quantum bus to nearby $^{13}$C nuclear spins \cite{Neumann2008, Childress2006, Neumann2010}. We use this configuration because the electron spin interacts strongly with control fields but is a poor qubit due to its short coherence time. On the other hand, the nuclear spins have a long coherence time but cannot be controlled easily due to their weak interactions \cite{Maurer_2012_RT}. One major limitation of this approach is that the NV$^-$ center's ground state electron spin, because its relaxation decoheres the nearby nuclear spins \cite{Filidou2012, Neumann2008}. Efforts have been made to mitigate this problem by electronically switching the NV into a spinless charge state \cite{Pfender2017}. However, this proves challenging from an engineering perspective.

The ST1 center promises to avoid this issue completely since it has a singlet (spinless) electronic ground state, which allows long nuclear spin coherence, and a photo-excited metastable triplet level, which can realize a quantum bus. The ST1 center also has brighter fluorescence and optical spin-readout contrast than the NV center, making it a strong candidate for the next generation of hybrid diamond quantum computing technologies \cite{lee_2013_ST1,L2_center_2017}.
 
However, progress in utilizing the ST1 center for quantum technology has been frustrated because it has been difficult to reproduce. Indeed, detection of the ST1 center has only been reported twice: first by \citet{lee_2013_ST1} in ultra-pure single crystalline HPHT diamond after fabrication of vertical nanowires and later, by \citet{L2_center_2017} in ion-implanted single crystal CVD diamond. Since both observations were in manufactured diamond, ST1 creation has been attributed to the synthesis process and the long-term stability of the defect is unknown. Due to the scarcity of samples, there are large gaps in our understanding of the ST1. Principally, its chemical composition and structure is unknown. Its electronic structure has been partially identified, but it may still have additional levels. There has been no work on studying the ST1 center's electron-phonon interactions or ensemble variation. 

In this paper we report the first observation of ST1 centers in natural diamond, indicating defect stability on geological timescales. We also conducted optical characterization of these ST1 that revealed unknown electronic levels; pump power dependent photodynamics that demonstrate an impressive (up to 80\%) optical spin readout contrast at high laser power; and new insight into its electron-phonon interactions which we use to identify the possible chemical structures of the center. These observations provide significantly more insight into the nature of the center and, by shortlisting possible structures, our analysis will enable future studies to be more targeted. This will lead more rapidly to precise identification of the defect. 

This paper has the following structure. In \cref{sec:theST1Center} we present an overview of the known characteristics of the ST1. In \cref{sec:experimentalDetails} we detail the experimental apparatus we used to study our sample. In \cref{sec:Results and Analysis} we present the optical characterization of the defects found in the sample, including a study of the site-to-site variation of its zero phonon line (ZPL). We then present the photodynamics experiments which demonstrate additional levels in its electronic structure and optical properties. In \cref{sec:PSBdecomp} we decompose our high-resolution optical spectra to obtain the electron-phonon spectral density. We use this spectral density and critical point analysis in \cref{sec:idthedefect} to identify a set of the simplest possible defect structures that are consistent with experiment.

\section{\label{sec:theST1Center} The ST1 Center}

A few of the key characteristics of the ST1 center are known to date. \citet{lee_2013_ST1} and \citet{L2_center_2017} both report that center's key optical features are a sharp ZPL around \SI{550}{nm} (\SI{2.25}{eV}) and broad phonon side-band (PSB) extending out to \SI{750}{nm}. These features indicate that the optical transition occurs between two discrete levels deep in the diamond bandgap. Their optical characterization established that it is a single emitter by observing the characteristic dip in photon-autocorrelation at zero-delay. The autocorrelation also showed pronounced bunching-shoulders which strongly indicated the presence of a long-lived shelving state.

The proposed electronic structure of the ST1 consists of singlet ground and excited states. The shelving state was determined to be a triplet. This was demonstrated by observing increased fluorescence corresponding to microwave fields resonant with three electron spin resonance transitions. The increase in fluorescence was explained by studying the optical dynamics of the system. The key features of the optical cycle are a rapid intersystem crossing (ISC) from the excited to the shelving state followed by another ISC to the ground state. The decay rate of the lower ISC is determined by the lifetimes of the triplet sublevels: $\tau_{\ket{0}}\approx\SI{2500}{ns}$, $\tau_{\ket{-1}}\approx\SI{1000}{ns}$ and $\tau_{\ket{+1}}\approx\SI{250}{ns}$.
Since the $\ket{0}$ state is much longer lived than the other sublevels, the system can be spin polarized into this sublevel. Due to the substantial differences in sub-level lifetimes, the rate of non-radiative decay to ground state via the metastable state can be significantly enhanced when spin population is driven out of the longer-lived triplet sub-levels via resonant microwave excitation. This results in an increase in average fluorescence, thereby enabling spin readout and the optical detection of magnetic resonance. The read out contrast is up to 45\%.

The optically detected magnetic resonance (ODMR) spectra of the of the triplets showed fine-structure at zero field: $2E = \SI{278(1)}{MHz}$, $D-E=\SI{996(1)}{MHz}$, and $D+E=\SI{1274(1)}{MHz}$. The spin-Hamiltonian of the triplet manifold is
\begin{equation}
     \hat H = D\bqty{S_z^2 -S(S+1)/3} + E \pqty{S_x^2 - S_y^2} + \gamma_e \vec S\cdot \vec B
     \label{eq:hamiltonian}
\end{equation}
where $\vec S = \pqty{S_x,S_y,S_z}$ are the usual dimensionless spin operators, $S=1$ is total spin, $\vec B$ is an external magnetic field, $\gamma_e$ is the free-electron gyromagnetic ratio (the observed g-factor is $g\approx2.0(1)$) and both $D=\SI{1135(1)}{MHz}$ and $E=\SI{139(1)}{MHz}$ are the zero field splitting parameters. Rotation of an applied magnetic field demonstrated that the spin-quantization axis of the triplet in the $\expval{110}$ crystalline direction, this defines the orientation of the spin-operators in \cref{eq:hamiltonian} such that $z||\expval{110}$.
The presence of the $E$ parameter suggests the center also has a minor spin axis in an orthogonal direction to $\expval{110}$ (e.g. $x||\bqty{100}$, $y||\bqty{011}$ and $z||\bqty{01\bar{1}}$) and that the center has C$_{2v}$ or lower symmetry. Optical polarization studies of the defect show that the transition dipole moment of the main transition is also in $\expval{110}$. However, it is unknown if the spin-quantization axis and dipole moment are co-aligned. Hyperfine splitting of the triplet due to nearby $^{13}$C was observed  but no further hyperfine structure intrinsic to the defect was identified. This seemingly suggests that the most prevalent isotopes of the center's chemical constituents are spinless.
\begin{figure}
\centering
\includegraphics[width=0.4\textwidth]{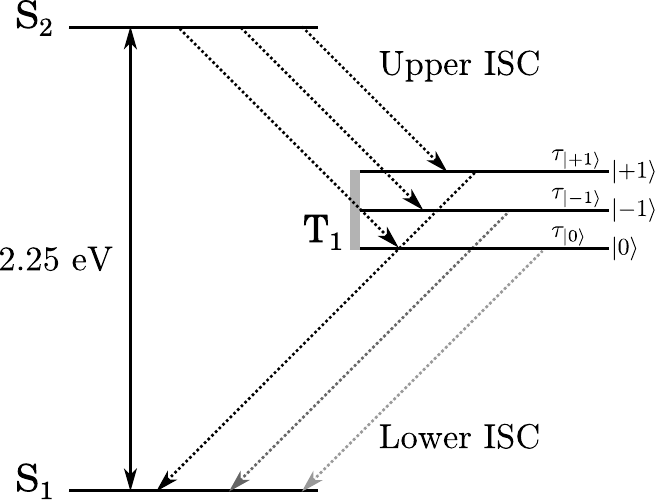}
\caption{\label{fig:ElectronicStructure} 
The known electronic structure of the ST1 center. $S_1$ and $S_2$ are the ground and excited singlets, respectively. $T_1$ is the metastable triplet state. The dotted lines indicate the ISCs and the solid arrow represents the optical ZPL. The lower ISCs are labelled by their depopulation rates, $\tau_{\ket{0}}$, $\tau_{\ket{+1}}$ and $\tau_{\ket{-1}}$.}
\end{figure}

\section{\label{sec:experimentalDetails} Experimental Details}
Our experimental procedure involved identification and analysis of ST1 centers in a natural diamond sample which was host to a wide verity of fluorescent sites. Our primary tools were optical microscopy, optically detected magnetic resonance (ODMR), optical spectroscopy and optical dynamics.

For our optical characterization we employed a home-built confocal fluorescence microscope with a \SI{532}{nm} green excitation laser. The fluorescence was  detected  using  either  an  avalanche  photodiode (APD) or a spectrometer. For our ODMR measurements, we applied microwaves through a copper wire positioned above the sample and an external magnetic field using a permanent magnet. All of our results were obtained at room temperature except our emission spectra which were obtained at 5K. For our optical polarization studies, a half wave plate (HWP) along with a linear polarizer was placed in the detection path. Polarization dependence was measured by rotating the HWP.
Our study of the optical dynamics of the system analyzed the  
second order photon correlation function, $g^{(2)}$ at different excitation powers. The autocorrelation function from single sites was measured in a Hanbury Brown and Twiss configuration, where the fluorescence was split by a 50:50 beamsplitter and detected using two APDs.

\section{\label{sec:Results and Analysis} Experimental Results}

\begin{figure}
\centering
\includegraphics[width=0.45\textwidth]{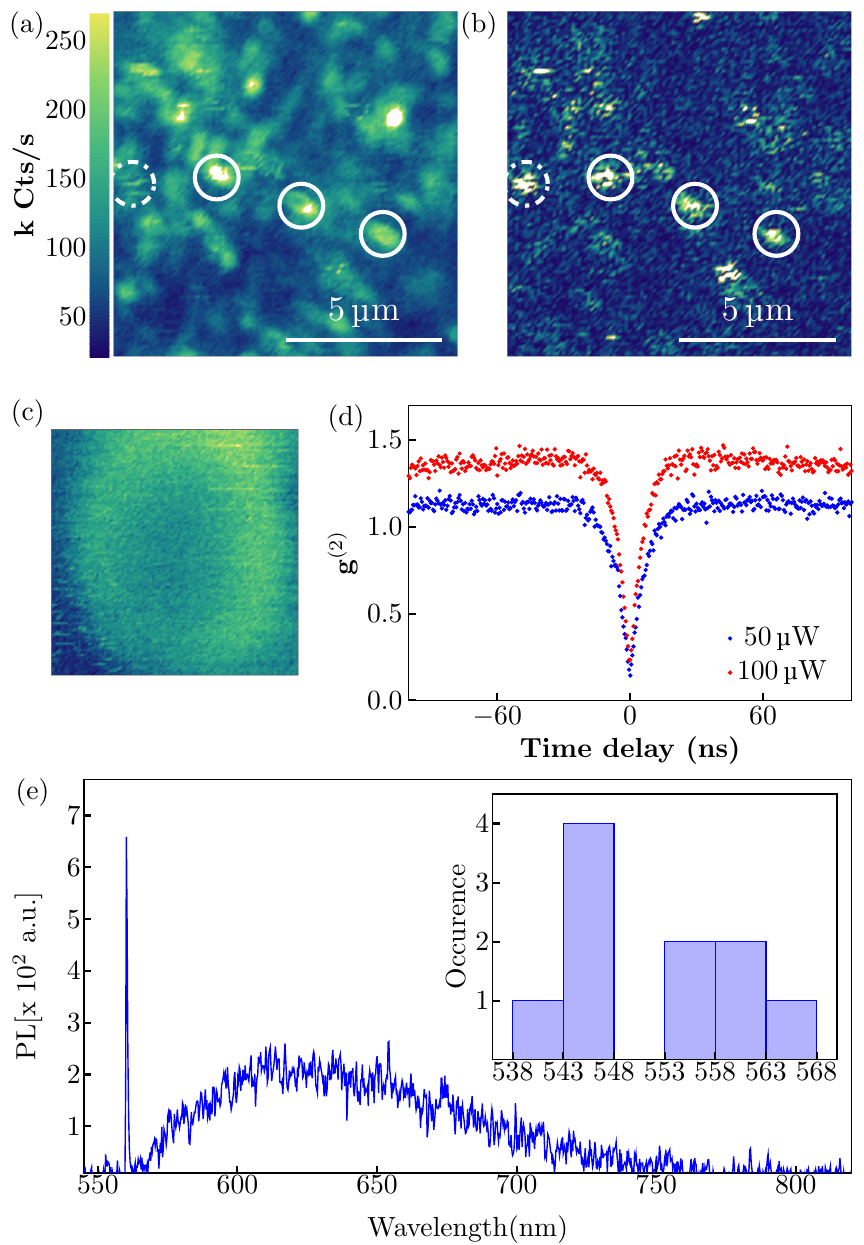}
\caption{\label{fig:confocal_1} (a) A confocal fluorescence image of the natural diamond sample. The ST1 centers are identified by circles. A solid circle indicates that the center can be spatially isolated from other sources. (b) A confocal fluorescence image of the same region shown in \cref{fig:confocal_1} (a) with applied resonant microwaves. (c) Magnified image of a single site at high laser power, showing a doughnut shaped fluorescence pattern. (d) Normalized photon autocorrelation function measured at different laser power showing anti-bunching at $ \tau =0 $. (e) The emission spectra of ST1 at 5K with a prominent ZPL at 555nm and PSB extending to 700nm. The inset depicts a histogram showing the distribution in the position of the ZPL.
}\end{figure}

\begin{figure}[h!]
\centering
\includegraphics[width=0.45\textwidth]{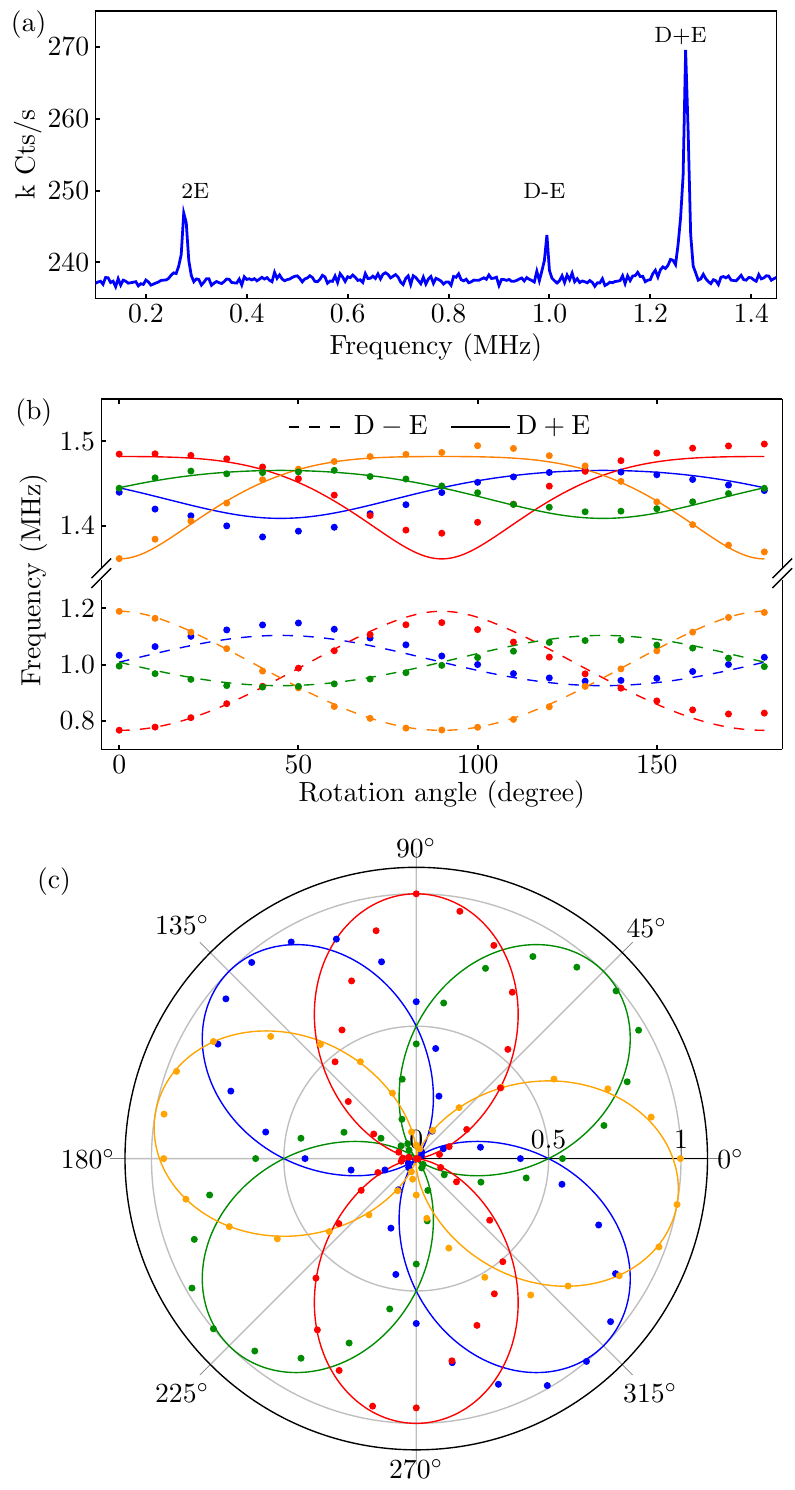}
\caption{\label{fig:magnetic_dipole_orientation}(a) ODMR spectrum of ST1 at zero magnetic field. (b) The response of the ODMR frequencies when a $B=\SI{120}{Gauss}$ magnetic field is rotated in the [001] crystal plane. The solid lines represent a fit using \cref{eq:hamiltonian}. (c) Plot of the polarization dependent fluorescence of the ST1 centers. It shows four linear polarization states, defined as $\phi_1=0\degree$, $\phi_2=45\degree$, $\phi_3=90\degree$ and $\phi_2=135\degree$.}
\end{figure}

We started by scanning a natural diamond sample using our confocal microscope. As shown in \cref{fig:confocal_1}~(a) we find a host of fluorescent sites in the sample. Of these, we observed that many sites increased in fluorescence when we applied microwaves resonant to the known ST1 spin transitions. The sites, marked in \cref{fig:confocal_1} (b), were abundant and uniformly distributed throughout the sample. We identified them as ST1 centers from their emission spectra, depicted in \cref{fig:confocal_1} (e), and ODMR signature, shown in \cref{fig:magnetic_dipole_orientation} (a). We repeated magnetic field rotation, \cref{fig:magnetic_dipole_orientation} (b), and polarometry, \cref{fig:magnetic_dipole_orientation} (c), to show that the centers have their major spin axis and dipole moment oriented in the $\expval{110}$  directions. We note that the fit of the angular dependence of the ODMR lines in \cref{fig:magnetic_dipole_orientation} (b) does not perfectly match the data when the field is aligned with $\expval{110}$. This may indicate that the major axis is slightly misaligned from $\expval{110}$ direction.
The $g^{(2)}$ autocorrelation for the detected sites, \cref{fig:confocal_1}~(d), shows a dip at zero delay and has bunching shoulders at higher excitation powers, which is consistent with the previous observations.
We did not observe any hyperfine structure intrinsic to the defect. This supports previous suggestions that the center's chemical constituents are spinless.

\begin{figure*}
\includegraphics[width=\textwidth]{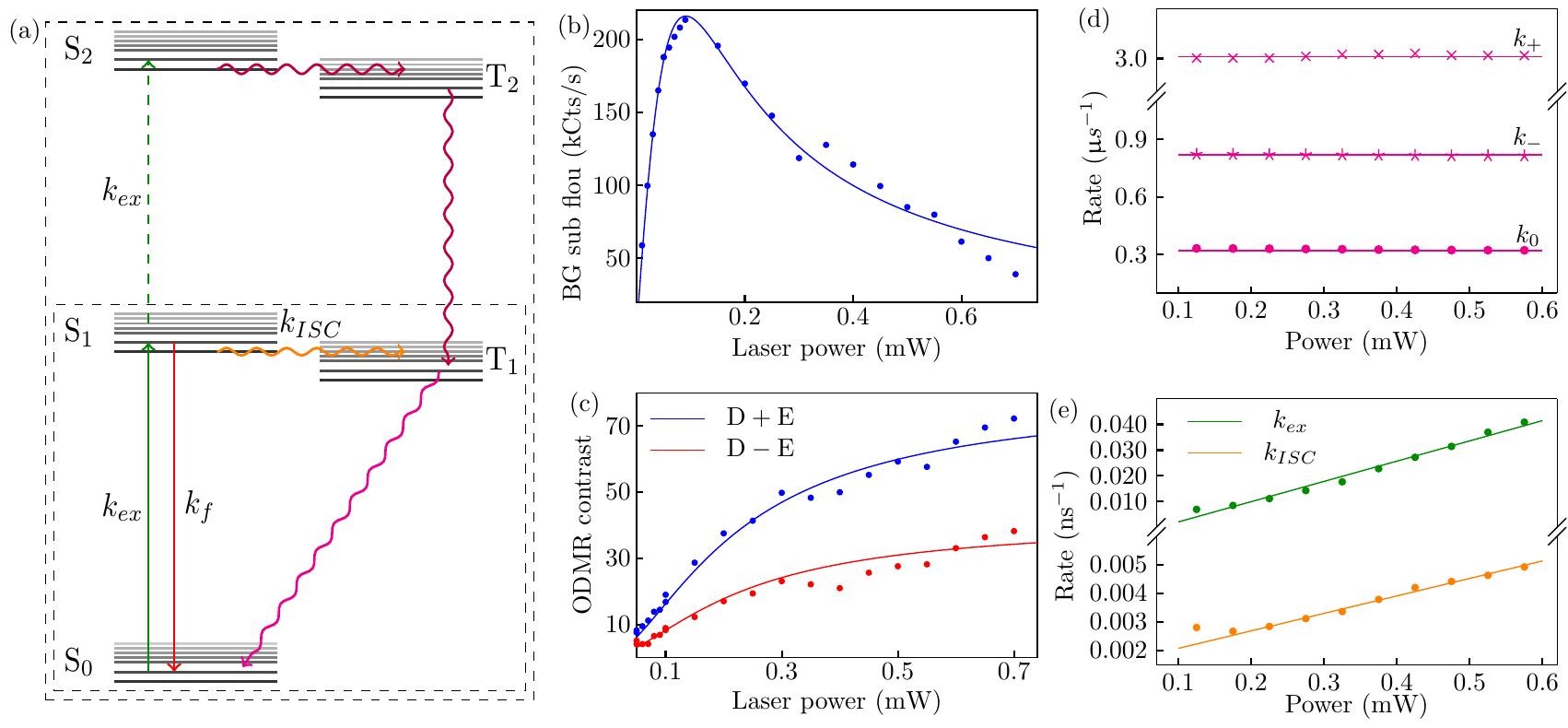}
\caption{\label{fig:fluorescence} (a) A proposed level scheme which could explain the power dependence of the optical dynamics. It includes a second excited singlet, $S_2$, and associated triplet state, $T_2$. The remaining sub-figures show the optical dynamics as a function of pump power. The solid lines represent a fit using a the rate equation model in \cref{eq:rateeq}. (b) depicts fluorescence counts. The solid lines represent the fit using the extended rate equation. (c) shows ODMR contrast. (d) is the depopulation rates from the triplet levels. (e) shows optical excitation rate $k_{ex}$ (green) and the upper intersystem crossing rate $k_{ISC}$ (orange).}
\end{figure*}

We identified ten sites and studied the variation between ST1 centers from site to site. The spread of the ZPL wavelengths in the sample of centers we measured was $\approx\SI{20}{nm}$; the ZPL frequency distribution is inset as a histogram in \cref{fig:confocal_1} (d). Notably, each of the defects measured showed similar ODMR spectra. Since the ZPL is expected to be significantly more susceptible to strain than the ODMR features, we attribute the distribution of ZPL energies to local variation of strain between sites.

A new feature we observed was an interesting power dependence of the center's photoluminescence. Between laser powers of $\approx$ \SI{100}{\micro\watt} and \SI{200}{\micro\watt}, the defect did not show blinking and had a brightness comparable to that of the NV centers. However, as we swept to higher laser powers (up to \SI{800}{\micro\watt}), the defects exhibited peculiar saturation behavior. This manifested in a doughnut shaped fluorescence pattern in our confocal scans, shown in \cref{fig:confocal_1} (c). The dip in fluorescence is due to the gaussian profile of the laser beam, which only saturates fluorescence as the center of the laser spot passes over the defect. This was further characterized in \cref{fig:fluorescence}, where we measured total fluorescence while sweeping laser power.
The NV center shows a similar fluorescence dip due to photoionization into its neutral (NV$^0$) charge state. However, unlike the NV, high power does not result in a reduction of the ST1 center's ODMR contrast. This excludes the possibility of defect photoionization. In fact, as demonstrated in \cref{fig:fluorescence} (c), increasing laser power leads to a impressively higher ODMR contrast of up to $\approx 80\%$. This motivated us to revisit of the photodynamics of the ST1 center.

We started with the simplest rate equation model that describes the established electronic structure. This is given by 
\begin{equation}
    \label{eq:rateeq}  
    \begin{split}  
     \dot{\text{S}}_\text{0} &= -k_{ex} \text{S}_\text{0} + k_f \text{S}_\text{1} + k_0 \text{T}_0 + k_- \text{T}_- + k_+ \text{T}_+ \\
     \dot{\text{S}}_\text{1} &= k_{ex} \text{S}_\text{0} - k_f \text{S}_\text{1} - k_{ISC} \text{S}_\text{1} \\
     \dot{\text{T}}_+ &= \frac{k_{ISC}}{3} \text{S}_\text{1} - k_+ \text{T}_+ \\
     \dot{\text{T}}_- &= \frac{k_{ISC}}{3} \text{S}_\text{1} - k_- \text{T}_- \\
     \dot{\text{T}}_0 &= \frac{k_{ISC}}{3} \text{S}_\text{1} - k_0 \text{T}_0 
     \end{split}
\end{equation}
where excitation from the ground to the excited sate is parameterized by $k_{ex}$, $k_f$ is the excited state's fluorescent lifetime and the parameters $k_+$, $k_-$ and $k_0$ are the depopulation rates from the respective triplet sublevels. For simplicity, we ignore the differences in the upper intersystem crossings for each triplet, assuming that they are all $k_{ISC}$. These rates are derived by solving the rate equation and fitting the $g^{(2)}$ autocorrelation, we provide details of this procedure in Appendix~S1.

We interrogated the power dependence of the rate equation parameters by fitting them at different laser powers. \cref{fig:fluorescence} (d) shows that the depopulation rate from the triplet sublevels is independent of the pump power. This eliminates the possibility of a intensity dependent depopulation channel from the triplet sublevels to the ground state. \cref{fig:fluorescence} (e) depicts the power dependence of $k_{ex}$ (green). As expected $k_{ex}$ increases linearly with pump power, as described by the equation 
\begin{align}
	\label{eq:kexvsP}
     k_{ex} = \frac{\lambda}{hc}\sigma(\lambda) I
\end{align}
where $I$ is the applied focal irradiance, $\sigma(\lambda)$ is the corresponding absorption cross-section at excitation wavelength $\lambda$. By fitting the slope of $k_{ex}$ we can extract an absorption cross-section of $\sigma \approx 10^{-17}~\mathrm{cm}^2$ which is comparable to the reported absorption cross-section of the NV center. Fig.~\ref{fig:fluorescence} (e) also shows an unexpected linear dependence of $k_{ISC}$ on pump power (orange). Since the upper intersystem crossing rate should be independent of pump power, it suggests that there is another channel for population transfer from the excited state to the triplet. A simple explanation for this observation is the existence of a second set of excited singlets and triplets (as per \cref{fig:fluorescence}). By assuming that the lifetimes of the higher excited states are very small, we can include them phenomenologically into our model (described in Appendix~S1) to estimate the second excitation rate. From this, we find a second absorption cross-section of $\sigma \approx 10^{-18} \mathrm{cm}^2$.

\section{\label{sec:PSBdecomp}Decomposition of the Phonon Side Band}
Focusing now on a detailed analysis of the optical band. The phonon-sideband (PSB) of an optical transition is generated by the electron-phonon interactions involved in the transition. In the absence of vibronic interactions, such as the Jahn-Teller effect, the PSB can be described by the linear symmetric mode model \cite{Maradudin_Point_Defects}. The linear symmetric mode model was applied extensively by \citet{Davies_Diamond} in the analysis of diamond color centers. Davies showed how the model could be used to extract the electron-phonon spectral density, referred to as the one-phonon band. Applying critical-point analysis allows this spectral density to be interpreted in terms of structural features of the defect \cite{Kehayias_Infrared_Spectra}. This is the approach adopted in this section.

Given an emission band $I_{em}(\omega)$, the bandshape function, $I\pqty{\omega}$, of a center is proportional to $I_\text{em}\pqty{\omega-\omega_0} \omega^{-3}$, where $\omega_0$ is the zero phonon line (ZPL) frequency and $I_\text{em}$ is the emission spectra. The band function is given by
\begin{align}
\label{eq:PSBeqn}
I(\omega) = e^{-S} I_0(\omega)\otimes\bqty{\delta(\omega)+\sum_{n=1}^\infty \frac{S^n}{n!} I_n(\omega)}
\end{align}
where $S$ is the total Huang-Rhys factor, $I_\text{0}$ is the ZPL shape, and $\otimes$ denotes convolution. The function
\begin{equation}
        I_n\pqty{\omega}=\int^\infty_{-\infty}I_{n-1}\pqty{\omega-x}I_1\pqty{x}dx
\end{equation}
is the $n$-phonon band which is constructed by successive convolutions of the one-phonon band. The one-phonon band represents all processes involving the creation and annihilation of a single phonon. Through self-convolutions, it generates the $n$-phonon band $I_n$ which describes all $n$-phonon processes with total energy $\hbar \omega$.

In theory, one can directly extract the one-phonon band from the band function by applying an inverse Fourier transform to \eqref{eq:PSBeqn}, rearranging to obtain an expression for $I_1(t)$, and then applying a Fourier transform,  $\hat{\mathcal{F}}$, to obtain
\begin{equation}
		\label{eq:fourierdeconv}
		I_1(\omega) = \frac{1}{S} \hat{\mathcal{F}} \bqty{ \log\pqty{\hat{\mathcal{F}}^{-1} \bqty{ I\pqty{\omega} - e^{-S} I_0\pqty{\omega} }} }
\end{equation}

However, due to the largely featureless PSB of the ST1\@ center, the large Huang-Rhys factor, and the comparatively low signal-to-noise of our experimental spectra, this direct Fourier deconvolution method is difficult because it is sensitive to numerical and spectral noise. Thus, the direct method is only sufficient to obtain an initial estimate of $I_1(\omega)$.

\begin{figure}[h!]
    \centering
    \includegraphics[width=0.43\textwidth]{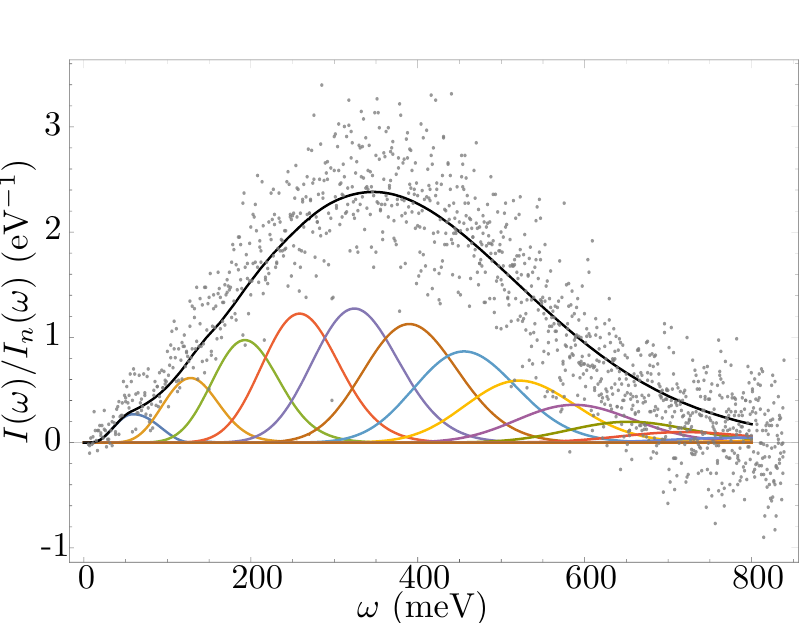}
 	\caption{The PSB spectrum from experiment (gray points), a polynomial fit of the spectrum to show its center-line (dashed blue), the calculated PSB (dashed orange), and its constituent n-phonon bands. The sum of n-phonon bands equals the calculated PSB. }
     \label{fig:nconvolutions}
\end{figure}

These issues can be overcome by using an iterative deconvolution method developed by. We applied this method by first obtaining an approximate one-phonon band from Fourier deconvolution. We then smoothed and tapered this approximate one-phonon band to form our first estimate $I_1^0(\omega)$ that was appropriately continuous and restricted to $\omega\in\bqty{0,\Omega}$, where $\Omega$ is the phonon cut-off of diamond. Next, we calculated the normalized PSB components $I_n^{0}(\omega)$ via successive convolutions of $I_1^{0}(\omega)$. We applied the following equation to generate an improved estimate $I_1^1(\omega)$ for the one-phonon band
\begin{align}
I_1^k(\omega) = e^S I(\omega) - I_{0}(\omega) - \sum_{n=2}^\infty \frac{S^n}{n!} I_{0} \otimes I_n^{k-1} (\omega).
\end{align}
where $k$ is the inductive step index. We then inductively repeated this procedure of calculating the normalized PSB components and generating the next estimate of the one-phonon band until it converged. This processes only required a small number of iterations and the results are depicted in \cref{fig:nconvolutions}. As can be seen, the generated band function matches the central line of the spectrum very well.

\begin{figure*}
	\centering
	\includegraphics[width=\textwidth]{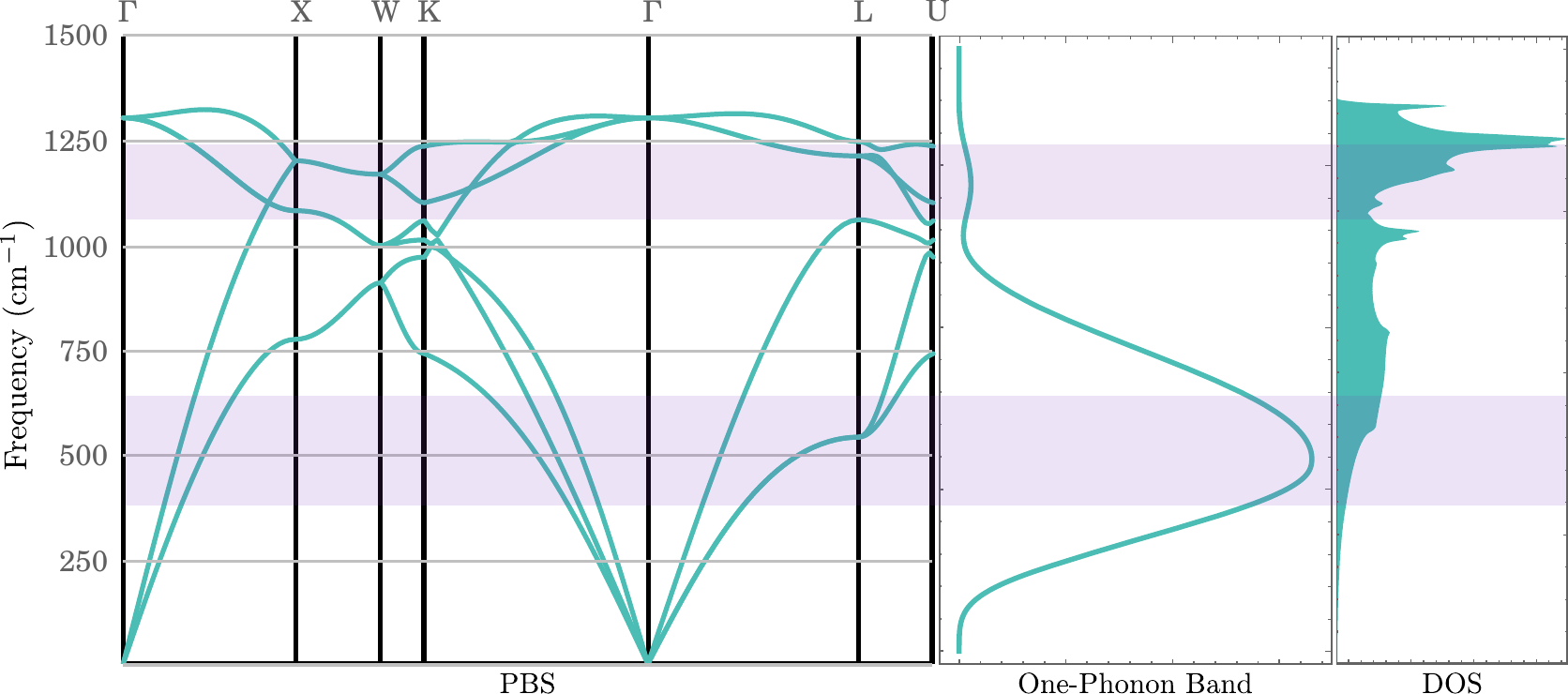}
	\caption{The comparison of the one-phonon band with the DOS and PBS of pristine diamond \cite{Petretto_DFT_BS}. The shaded areas connect adjacent points to highlight regions of interest.}
	\label{fig:bscomp}
\end{figure*}

We now turn to critical point analysis to relate the features of the one-phonon band to structural components of the defect. The key assumption of critical point analysis is that the center does not significantly perturb the phonon modes of pristine diamond. Using this approach features in the one-phonon band correspond to either frequencies of high mode density and/ or where there is strong coupling to the defect orbitals. \cref{fig:bscomp} shows the extracted one-phonon band against the phonon band structure (PBS) and density of states (DOS) of diamond. The absence of spectrally sharp features at frequencies above $\Omega$ shows that the optical transition does not couple to local modes; only weak coupling to the continuum modes is present. This validates the key assumption of our application of critical point analysis. 

As shown in \cref{fig:bscomp}, the one-phonon band's largest feature is a broad peak centered at 60 meV. The prominence of this feature indicates that at this frequency, two things are occurring: a high density of modes and strong coupling to the defect. Indeed, the feature is coincident with the ``leveling out'' of the transverse phonon bands at the $L$-point, which implies a higher relative density of modes of that phonon type. Furthermore, since the $L$-point lies on the edge of the Brillouin zone, these phonons also result in maximum displacement between equivalent atoms in neighboring unit cells. This implies that for the defect to strongly couple to these modes, its orbitals must be well-localized to the nearest-neighbor atoms of a lattice site. This is just like the orbitals of the NV$^-$ center in diamond that surround a vacant lattice site.

\begin{figure}[h!]
	\centering
	\includegraphics[width=0.43\textwidth]{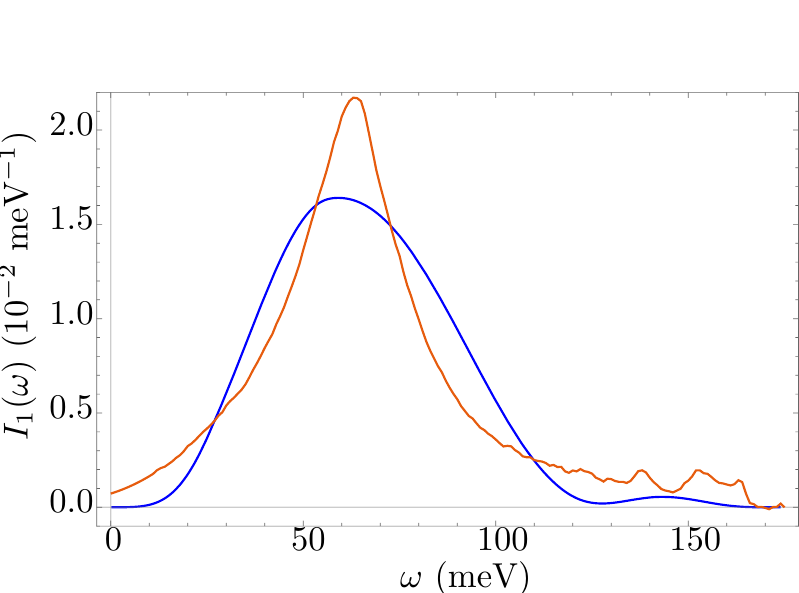}
	\caption{Comparison of the one-phonon bands of the NV$^{-1 }$\cite{Kehayias_Infrared_Spectra} and ST1 optical transitions.}
    \label{fig:NVcompare}
\end{figure}

\cref{fig:NVcompare} shows that the one-phonon bands of ST1\@ and NV$^-$ are indeed remarkably similar. Assessment of the critical points of the NV$^-$ phonon band by \citet{Kehayias_Infrared_Spectra} showed that it also couples most strongly to phonon modes at the $L$-point. They observed that $L$-point modes would result in the greatest distortion of the electron density localized to the dangling $sp^3$ orbitals about the center's vacancy. The strong similarities of the one-phonon bands of ST1\@ and NV$^-$ strongly indicate that the ST1 center contains a vacancy and the orbitals involved its optical transition are highly localized to this vaidthedefectcancy. We use this conclusion to greatly simplify the identification of possible defect structures of the ST1 center.

\section{\label{sec:idthedefect}Identifying the ST1\@}

\begin{figure*}[h!]
    \centering
	\includegraphics[width=0.9\textwidth]{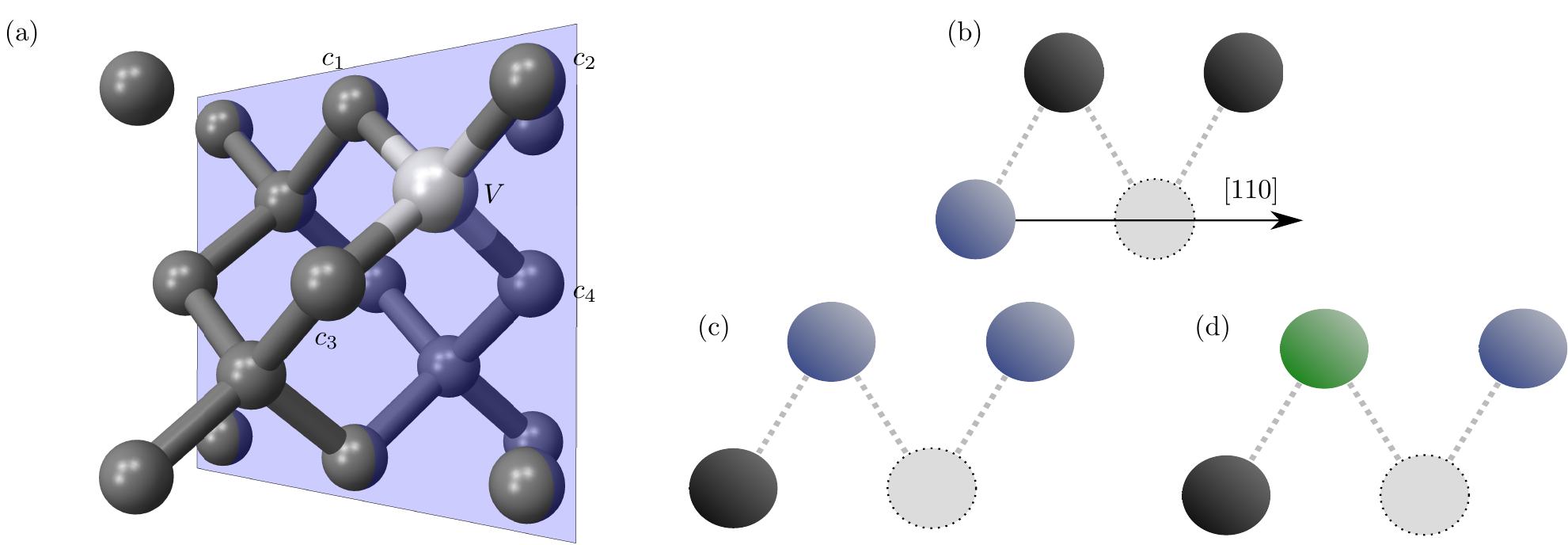}
    \caption{(a) is a cell of diamond showing a vacancy and the [110] reflection plane. The other figures are the different possible geometries represented by a [110] cross-section of the defect. If we let X and Y represent potential impurities (b) is a next-to-nearest-neighbour substitution, labelled XCV; (c) is the nearest-neighbour substitutions of like atoms, labelled XVX; and (d) is nearest-neighbour substitutions of unlike atoms, labelled XVY.}
	\label{fig:drawing}
\end{figure*}

In this section, we outline the simplest defect structures that are consistent with the experimental understanding of the ST1\@ center acquired to date. The analysis of the PSB motivates studying vacancy centric models for the ST1\@. Recapping, the center has $C_{2v}$ symmetry or lower and it is oriented along the [110] axis. These two pieces of information constrain the simplest defect structures to those containing a vacancy with nearby substitutional lattice impurities orientated along the [110] axis. We accordingly restrict the geometries considered in this work to substitutions of either the nearest neighbors to the vacancy (NNVs) or of its next-to-nearest neighbors within the same reflection plane of the defect. Schematics of these geometries are provided in \cref{fig:drawing}. 

The shortlisted structures can be further constrained by considering that that intrinsic hyperfine structure has not been detected for the ST1 center. Since the electron spin-density of the defect is localized to the vacancy's nearest-neighbour atoms, the absence of hyperfine structure particularly constrains defects with substitutions of these atoms to those without nuclear spin. Alternatively, a lack of hyperfine would support a next-to-nearest neighbor substitution as spin-density would be localized to the vacancy and, thus,  unlikely to couple strongly to a nuclear spin two atomic sites distant. 

Since the ST1\@ has a ground singlet, it has integer spin. This means that the defect consists of an even number of electrons. The NNVs contribute at least a total of four electrons to the defect. Impurities may only contribute total of two additional electrons or none. The defect therefore has either four or six electrons. Ignoring positive charge states due to their low prevalence in diamond, we propose the following candidate impurities: [$Si$], [$N$]+e$^-$, and [$O$] (where [] indicates that it could be any species in the column of the periodic table). We list a selection of the simplest possible ST1\@ geometries in \cref{tab:ST1geom}.

\begin{table}[]
\centering
\begingroup
\setlength{\tabcolsep}{10pt} 
\renewcommand{\arraystretch}{1.5} 
\begin{tabular}{c | c c c}
 & \multicolumn{3}{c}{Defect Symmetry and Structure} \\
e No. & $C_{1h}$ & $C_{1h}$ & $C_{2v}$ \\
 \hline
$4$ & $[Si]CV$ & - & $[Si]V[Si]$ \\
$6$ & $[O]CV$, $[N]^-$ & $[O]V[Si]$ & - \\
\end{tabular}
\endgroup
\caption{Our selection of the simplest possible candidates for the ST1. The labelling convention follows the one defined in \cref{fig:drawing}}
\label{tab:ST1geom}
\end{table}

Having identified the set of simplest defect structures, we will now construct their electronic structures to test if they are consistent with experiment. We do this by applying the standard defect molecule model \cite{Loubser_1978_ESRdiamond}. Adopting a minimal basis of the four dangling sp$^3$ orbitals from the atoms surrounding the vacancy, we combine these linearly to form a basis of symmetrized molecular orbitals. Considering a defect with $C_{2v}$ symmetry, we form the following molecular orbitals 
\begin{align}
A_1: a_{1} &= c_1 + c_2 \nonumber \\
A_2: a_{1}' &= c_3 + c_4 \nonumber \\
B_1: b_{1} &= c_1 - c_2 \nonumber \\
B_2: b_{2} &= c_3 - c_4
\end{align}
where $c_1$ and $c_2$ are the orbitals centered on the nearest-neighbours atoms in the (110) plane, while $c_3$ and $c_4$ are the out-of-plane atoms. $A_1$, $A_2$, $B_1$ and $B_2$ are the irreducible representations of $C_{2v}$. Notably, these orbitals are the same as those for a $C_{1h}$ defect. In this case, the symmetry of the $A_1$ and $B_1$ states is lowered to $A'$. While  $A_2$ and $B_2$ are lowered to $A''$.

\begin{figure*}[h!]
    \centering
    \def\svgwidth{\textwidth}
	\includegraphics[width=0.9\textwidth]{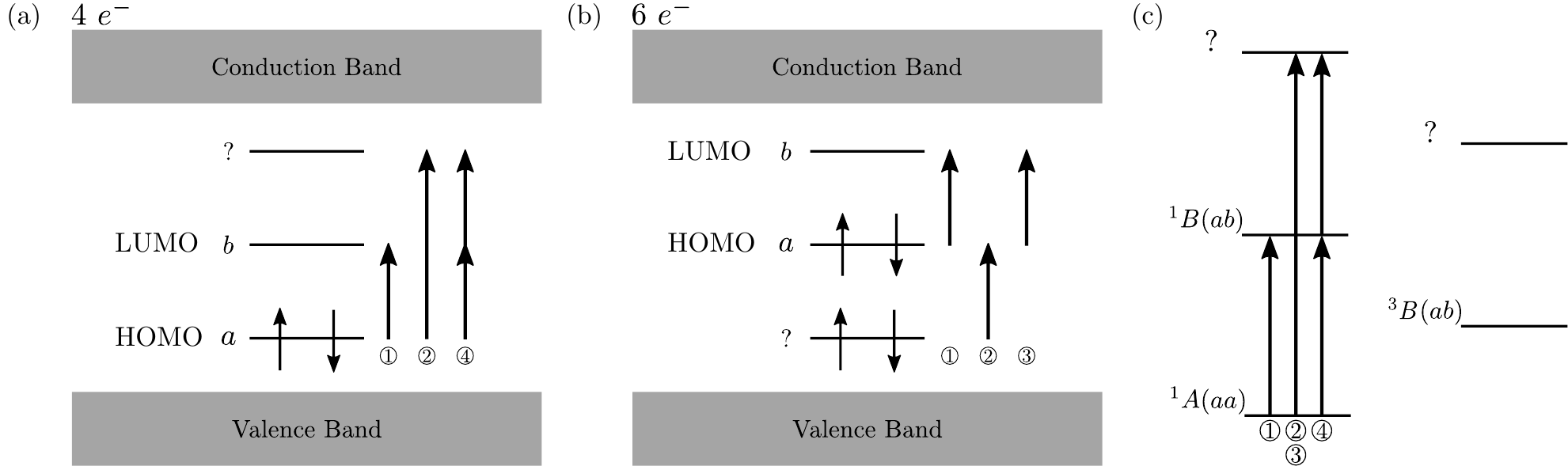}
    \caption{For an arbitrary HOMO/LUMO pair $(a,b)$, (a) and (b) are the proposed orbital configurations associated to the four and six electron molecular models, (c) is the associated electronic structure. The numbered arrows represent the allowed transition pathways to the first and second excited transitions. These, when combined with the orientation of the dipole moment, help identify the symmetry of the electronic states.}
	\label{fig:ES}
\end{figure*}

To proceed with constructing the electronic structure, we must now energetically order the MOs. The possible ordering options can be reduced by interpreting the orientations of the center's optical dipole moments and spin quantization axes. The absence of orbital degeneracy in $C_{2v}$ and lower symmetry, combined with the singlet ground state, implies that the highest occupied molecular orbital (HOMO) of the ground configuration is doubly occupied.  Hence, we can immediately identify the ground electronic state as a $^1A_1$ via {\"U}nsold's theorem. It follows that the dipole moment of the primary optical transition and the spin-quantization axes are determined by the levels of the first excited configuration, where one electron has been promoted from the HOMO to the lowest unoccupied molecular orbital (LUMO) of the ground configuration. \citet{lee_2013_ST1} found that the both dipole moment and the spin-quantization axis are oriented in one of the equivalent [110] directions, but could not distinguish if they were co-aligned. In Appendix~S2, we evaluate the optical selection rules and the interpret elements of the spin-spin tensor to infer the orientation of the dipole moment and spin-quantization axis, respectively. We find that the only HOMO/LUMO pairs that are consistent with experiment are ($a_1$,$b_1$) and ($a_1'$,$b_2$). 

Next we use the centre's secondary transition to determine the energetic ordering of the remaining MOs. If the defect comprises of four electrons, the second excited state is where an electron in the HOMO is promoted to the second-LUMO. If it has six electrons, there are two potential second excited configurations: (1) where an electron is promoted from the second-HOMO to the HOMO (which is partially occupied in the first configuration), or (2) promotion of both the HOMO electrons to the LUMO. \cref{fig:ES} shows the MOs and identifies the configurations that could be associated to the excited states. As we did with the first excited state, the transition dipole moment of the second transition can be used to identify the symmetry of the second excited state. 

Thus, future experimental work should seek to determine the polarization of this second transition to further constrain the possible identity of the molecular orbitals and electronic levels.

Having heavily restricted the large number of defect candidates, it is now  feasible to conduct \textit{ab initio} calculations. The electronic structure, transition dipole moment, vertical transition energy, and spin-spin tensor can be extracted from these calculations and compared to the experimental observations and theory presented in our work.

\section{Conclusion}
In this paper, we report the first discovery of ST1 centers in a natural diamond sample. We present an experimental study of its optical spectra and dynamics, particularly, their dependence on excitation power. We used theoretical tools to identify its possible chemical and electronic configuration. Our key results are as follows.
     (1) ST1 centers have stability on geological timescales.
     (2) The centers do not show hyperfine structure intrinsic to the defect.
     (3) The defect has previously unidentified electronic structure. We propose that this is second excited singlet and triplet states.
     (4) The ST1 center's spin readout contrast can be enhanced up to 80\% by increasing pump power.
     (5) The absorption cross-section of the first transition is $10^{-17} \mathrm{cm}^2$.
     The defect couples most strongly to phonons at the $L$-point, like the NV center. This strongly indicates a vacancy centered defect.
     (6) A selection of candidate ST1 chemical structures that are consistent with experiment have been identified in table~S2.
     (7) The possible ST1 electronic structures that are consistent with experiment have been identified in \cref{sec:idthedefect}.
These observations are a significant advance in our fundamental understanding of the center. Our work provides the groundwork to make identification of the defect via \textit{ab initio} simulations or experiment feasible. Consequently, it is a significant step towards the practical realization of an ST1 quantum bus.



\bibliographystyle{unsrtnat}
\bibliography{refs}

\begin{thebibliography}{15}
\providecommand{\natexlab}[1]{#1}
\providecommand{\url}[1]{\texttt{#1}}
\expandafter\ifx\csname urlstyle\endcsname\relax
  \providecommand{\doi}[1]{doi: #1}\else
  \providecommand{\doi}{doi: \begingroup \urlstyle{rm}\Url}\fi

\bibitem[Kurizki et~al.(2015)Kurizki, Bertet, Kubo, M{\o}lmer, Petrosyan, Rabl,
  and Schmiedmayer]{kurizki2015quantum}
Gershon Kurizki, Patrice Bertet, Yuimaru Kubo, Klaus M{\o}lmer, David
  Petrosyan, Peter Rabl, and J{\"o}rg Schmiedmayer.
\newblock Quantum technologies with hybrid systems.
\newblock \emph{Proceedings of the National Academy of Sciences}, 112\penalty0
  (13):\penalty0 3866--3873, 2015.

\bibitem[Waldherr et~al.(2014)Waldherr, Wang, Zaiser, Jamali,
  Schulte-Herbrüggen, Abe, Ohshima, Isoya, Du, Neumann, and
  Wrachtrup]{Waldherr2014}
G.~Waldherr, Y.~Wang, S.~Zaiser, M.~Jamali, T.~Schulte-Herbrüggen, H.~Abe,
  T.~Ohshima, J.~Isoya, J.~F. Du, P.~Neumann, and J.~Wrachtrup.
\newblock Quantum error correction in a solid-state hybrid spin register.
\newblock \emph{Nature}, 506:\penalty0 204, February 2014.
\newblock URL \url{https://doi.org/10.1038/nature12919}.

\bibitem[Neumann et~al.(2008)Neumann, Mizuochi, Rempp, Hemmer, Watanabe,
  Yamasaki, Jacques, Gaebel, Jelezko, and Wrachtrup]{Neumann2008}
P.~Neumann, N.~Mizuochi, F.~Rempp, P.~Hemmer, H.~Watanabe, S.~Yamasaki,
  V.~Jacques, T.~Gaebel, F.~Jelezko, and J.~Wrachtrup.
\newblock Multipartite entanglement among single spins in diamond.
\newblock \emph{Science}, 320\penalty0 (5881):\penalty0 1326--1329, 2008.
\newblock ISSN 0036-8075.
\newblock \doi{10.1126/science.1157233}.
\newblock URL \url{http://science.sciencemag.org/content/320/5881/1326}.

\bibitem[Childress et~al.(2006)Childress, Gurudev~Dutt, Taylor, Zibrov,
  Jelezko, Wrachtrup, Hemmer, and Lukin]{Childress2006}
L.~Childress, M.~V. Gurudev~Dutt, J.~M. Taylor, A.~S. Zibrov, F.~Jelezko,
  J.~Wrachtrup, P.~R. Hemmer, and M.~D. Lukin.
\newblock Coherent dynamics of coupled electron and nuclear spin qubits in
  diamond.
\newblock \emph{Science}, 314\penalty0 (5797):\penalty0 281--285, 2006.
\newblock ISSN 0036-8075.
\newblock \doi{10.1126/science.1131871}.
\newblock URL \url{http://science.sciencemag.org/content/314/5797/281}.

\bibitem[Neumann et~al.(2010)Neumann, Beck, Steiner, Rempp, Fedder, Hemmer,
  Wrachtrup, and Jelezko]{Neumann2010}
Philipp Neumann, Johannes Beck, Matthias Steiner, Florian Rempp, Helmut Fedder,
  Philip~R. Hemmer, J{\"o}rg Wrachtrup, and Fedor Jelezko.
\newblock Single-shot readout of a single nuclear spin.
\newblock \emph{Science}, 329\penalty0 (5991):\penalty0 542--544, 2010.
\newblock ISSN 0036-8075.
\newblock \doi{10.1126/science.1189075}.
\newblock URL \url{http://science.sciencemag.org/content/329/5991/542}.

\bibitem[Maurer et~al.(2012)Maurer, Kucsko, Latta, Jiang, Yao, Bennett,
  Pastawski, Hunger, Chisholm, Markham, Twitchen, Cirac, and
  Lukin]{Maurer_2012_RT}
P.~C. Maurer, G.~Kucsko, C.~Latta, L.~Jiang, N.~Y. Yao, S.~D. Bennett,
  F.~Pastawski, D.~Hunger, N.~Chisholm, M.~Markham, D.~J. Twitchen, J.~I.
  Cirac, and M.~D. Lukin.
\newblock Room-temperature quantum bit memory exceeding one second.
\newblock \emph{Science}, 336\penalty0 (6086):\penalty0 1283--1286, 2012.
\newblock ISSN 0036-8075.
\newblock \doi{10.1126/science.1220513}.
\newblock URL \url{http://science.sciencemag.org/content/336/6086/1283}.

\bibitem[Filidou et~al.(2012)Filidou, Simmons, Karlen, Giustino, Anderson, and
  Morton]{Filidou2012}
Vasileia Filidou, Stephanie Simmons, Steven~D. Karlen, Feliciano Giustino,
  Harry~L. Anderson, and John J.~L. Morton.
\newblock Ultrafast entangling gates between nuclear spins using photoexcited
  triplet states.
\newblock \emph{Nature Physics}, 8:\penalty0 596, July 2012.
\newblock URL \url{https://doi.org/10.1038/nphys2353}.

\bibitem[Pfender et~al.(2017)Pfender, Aslam, Simon, Antonov, Thiering, Burk,
  Fávaro~de Oliveira, Denisenko, Fedder, Meijer, Garrido, Gali, Teraji, Isoya,
  Doherty, Alkauskas, Gallo, Grüneis, Neumann, and Wrachtrup]{Pfender2017}
Matthias Pfender, Nabeel Aslam, Patrick Simon, Denis Antonov, Gergő Thiering,
  Sina Burk, Felipe Fávaro~de Oliveira, Andrej Denisenko, Helmut Fedder, Jan
  Meijer, Jose~A. Garrido, Adam Gali, Tokuyuki Teraji, Junichi Isoya,
  Marcus~William Doherty, Audrius Alkauskas, Alejandro Gallo, Andreas Grüneis,
  Philipp Neumann, and Jörg Wrachtrup.
\newblock Protecting a diamond quantum memory by charge state control.
\newblock \emph{Nano Letters}, 17\penalty0 (10):\penalty0 5931--5937, 2017.
\newblock \doi{10.1021/acs.nanolett.7b01796}.
\newblock URL \url{https://doi.org/10.1021/acs.nanolett.7b01796}.
\newblock PMID: 28872881.

\bibitem[Lee et~al.(2013)Lee, Widmann, Rendler, Doherty, Babinec, Yang, Eyer,
  Siyushev, Hausmann, Loncar, Bodrog, Gali, Manson, Fedder, and
  Wrachtrup]{lee_2013_ST1}
Sang-Yun Lee, Matthias Widmann, Torsten Rendler, Marcus~W. Doherty, Thomas~M.
  Babinec, Sen Yang, Moritz Eyer, Petr Siyushev, Birgit J.~M. Hausmann, Marko
  Loncar, Zolt{\'a}n Bodrog, Adam Gali, Neil~B. Manson, Helmut Fedder, and
  J{\"o}rg Wrachtrup.
\newblock Readout and control of a single nuclear spin with a metastable
  electron spin ancilla.
\newblock \emph{Nature Nanotechnology}, 8\penalty0 (7):\penalty0 487--492, June
  2013.
\newblock ISSN 1748-3387, 1748-3395.
\newblock \doi{10.1038/nnano.2013.104}.
\newblock URL
  \url{http://www.nature.com/nnano/journal/v8/n7/fig_tab/nnano.2013.104_F5.html}.

\bibitem[John et~al.(2017)John, Lehnert, Mensing, Spemann, Pezzagna, and
  Meijer]{L2_center_2017}
Roger John, Jan Lehnert, Michael Mensing, Daniel Spemann, Sébastien Pezzagna,
  and Jan Meijer.
\newblock Bright optical centre in diamond with narrow, highly polarised and
  nearly phonon-free fluorescence at room temperature.
\newblock \emph{New Journal of Physics}, 19\penalty0 (5):\penalty0 053008,
  2017.
\newblock URL \url{http://stacks.iop.org/1367-2630/19/i=5/a=053008}.

\bibitem[Maradudin(1966)]{Maradudin_Point_Defects}
A.A. Maradudin.
\newblock Theoretical and experimental aspects of the effects of point defects
  and disorder on the vibrations of crystals.
\newblock volume~18 of \emph{Solid State Physics}, pages 273 -- 420. Academic
  Press, 1966.
\newblock \doi{https://doi.org/10.1016/S0081-1947(08)60350-1}.
\newblock URL
  \url{http://www.sciencedirect.com/science/article/pii/S0081194708603501}.

\bibitem[Davies(1974)]{Davies_Diamond}
G~Davies.
\newblock Vibronic spectra in diamond.
\newblock \emph{Journal of Physics C: Solid State Physics}, 7\penalty0
  (20):\penalty0 3797, 1974.
\newblock URL \url{http://stacks.iop.org/0022-3719/7/i=20/a=019}.

\bibitem[Kehayias et~al.(2013)Kehayias, Doherty, English, Fischer, Jarmola,
  Jensen, Leefer, Hemmer, Manson, and Budker]{Kehayias_Infrared_Spectra}
P.~Kehayias, M.~W. Doherty, D.~English, R.~Fischer, A.~Jarmola, K.~Jensen,
  N.~Leefer, P.~Hemmer, N.~B. Manson, and D.~Budker.
\newblock Infrared absorption band and vibronic structure of the
  nitrogen-vacancy center in diamond.
\newblock \emph{Phys. Rev. B}, 88:\penalty0 165202, Oct 2013.
\newblock \doi{10.1103/PhysRevB.88.165202}.
\newblock URL \url{https://link.aps.org/doi/10.1103/PhysRevB.88.165202}.

\bibitem[Petretto et~al.(2018)Petretto, Dwaraknath, P.C.~Miranda, Winston,
  Giantomassi, van Setten, Gonze, Persson, Hautier, and
  Rignanese]{Petretto_DFT_BS}
Guido Petretto, Shyam Dwaraknath, Henrique P.C.~Miranda, Donald Winston, Matteo
  Giantomassi, Michiel~J. van Setten, Xavier Gonze, Kristin~A. Persson,
  Geoffroy Hautier, and Gian-Marco Rignanese.
\newblock High-throughput density-functional perturbation theory phonons for
  inorganic materials.
\newblock \emph{Scientific Data}, 5:\penalty0 180065 EP --, May 2018.
\newblock URL \url{http://dx.doi.org/10.1038/sdata.2018.65}.

\bibitem[Loubser and van Wyk(1978)]{Loubser_1978_ESRdiamond}
J~H~N Loubser and J~A van Wyk.
\newblock Electron spin resonance in the study of diamond.
\newblock \emph{Reports on Progress in Physics}, 41\penalty0 (8):\penalty0
  1201, 1978.
\newblock URL \url{http://stacks.iop.org/0034-4885/41/i=8/a=002}.

\end{thebibliography}


\begin{thebibliography}{3}
\providecommand{\natexlab}[1]{#1}
\providecommand{\url}[1]{\texttt{#1}}
\expandafter\ifx\csname urlstyle\endcsname\relax
  \providecommand{\doi}[1]{doi: #1}\else
  \providecommand{\doi}{doi: \begingroup \urlstyle{rm}\Url}\fi

\bibitem[Nizovtsev et~al.(2003)Nizovtsev, Kilin, Jelezko, Popa, Gruber, Tietz,
  and Wrachtrup]{Nizovtsev2003}
A.~P. Nizovtsev, S.~Ya. Kilin, F.~Jelezko, I.~Popa, A.~Gruber, C.~Tietz, and
  J.~Wrachtrup.
\newblock Spin-selective low temperature spectroscopy on single molecules with
  a triplet-triplet optical transition: Application to the nv defect center in
  diamond.
\newblock \emph{Optics and Spectroscopy}, 94\penalty0 (6):\penalty0 848--858,
  Jun 2003.
\newblock ISSN 1562-6911.
\newblock \doi{10.1134/1.1586735}.
\newblock URL \url{https://doi.org/10.1134/1.1586735}.

\bibitem[Weisstein(2002)]{weisstein2002vieta}
Eric~W Weisstein.
\newblock Vieta's formulas.
\newblock 2002.

\bibitem[Doherty et~al.(2014)Doherty, Struzhkin, Simpson, McGuinness, Meng,
  Stacey, Karle, Hemley, Manson, Hollenberg, and
  Prawer]{Doherty_NV_Under_Pressure}
Marcus~W. Doherty, Viktor~V. Struzhkin, David~A. Simpson, Liam~P. McGuinness,
  Yufei Meng, Alastair Stacey, Timothy~J. Karle, Russell~J. Hemley, Neil~B.
  Manson, Lloyd C.~L. Hollenberg, and Steven Prawer.
\newblock Electronic properties and metrology applications of the diamond
  ${\mathrm{nv}}^{\ensuremath{-}}$ center under pressure.
\newblock \emph{Phys. Rev. Lett.}, 112:\penalty0 047601, Jan 2014.
\newblock \doi{10.1103/PhysRevLett.112.047601}.
\newblock URL \url{https://link.aps.org/doi/10.1103/PhysRevLett.112.047601}.

\end{thebibliography}



\end{document}


\title{Supplementary Information: Discovery of ST1 centers in natural diamond}

\maketitle

\externaldocument{./paper}

\section{Rate Equation Fitting Procedure}
\begin{figure}
\centering
\includegraphics{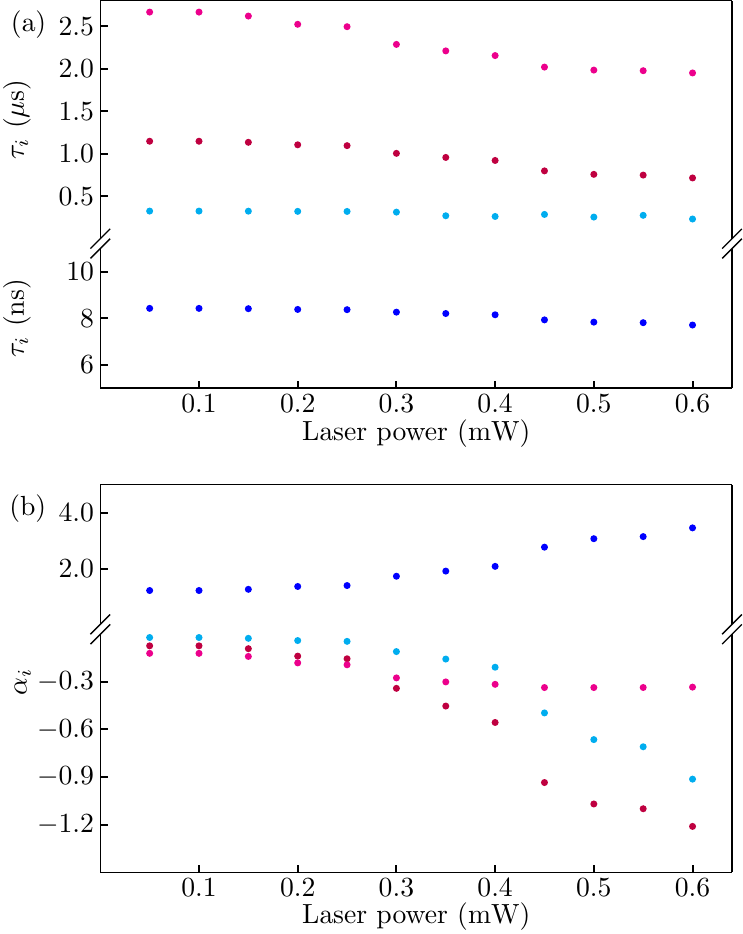}
\caption{\label{fig:decay_rates} Dependence of the $\tau_i$(a) and $\alpha_i$(b) 
 parameters as a function of the excitation power. Points are fits to the measured g(2) functions using \cref{eq:g2_fitfn}.
}\end{figure}

We investigated the photodynamics of the ST1 center using the second-order photon correlation, $g^{(2)}$, measured at different excitation powers. The recorded coincidence rate $c(t)$ is first normalized according to the formula $C_N(t)= c(t)/(N_1 N_2 w T)$, where $N_{1,2}$ are the counts on each APDs, $w$ is the bin width and $T$ is the total signal accumulation time. The $g^{(2)}(t)$ is obtained from the normalized coincidence rate $C_N(t)$ as $g^{(2)}(t) = (C_N(t)-(1-{\rho}^2))/{\rho}^2$, where $\rho$ is the signal to background ratio. The experimental results were fitted with the function
\begin{equation}
\begin{split}
 1-\sum_{i=1}^{4}\alpha_i\:e^{-t/\tau_i}
\end{split}
\label{eq:g2_fitfn}
\end{equation} 
where the $\alpha_i$ and $\tau_i$ are fit parameters. The power dependence of these parameters is shown in \cref{fig:decay_rates}.

\subsection{Five level system: Theoretical model}
The analytic form of the transient solution of the system described in eq.~(2) is found using Laplace transform. The autocorrelation function is written as \cite{Nizovtsev2003}
\begin{equation}\label{eq:auto_corr}
\begin{split}
\frac{\text{S}_{\text{1}}(\tau)}{\text{S}_{\text{1}}(\infty)} &=1-\sum_{i=1}^{4}e^{-\lambda_i t} \frac{\lambda_j\:\lambda_k\:\lambda_l}{k_0\:k_-\:k_+}\frac{\left(\lambda_i + k_0 \right)\left(\lambda_i + k_- \right)\left(\lambda_i + k_+ \right)}{(\lambda_i-\lambda_j) (\lambda_i-\lambda_k)(\lambda_i-\lambda_l)}
\end{split}
\end{equation}
where $ \lambda_i $ are the roots of the characteristic equation of the form $A x^4+B x^3+ C x^2+D x +E =0 $ with,
\begin{equation}\label{eq:poly_coeff}/
\begin{split}
A &=1 \\
B &= k_{ex}+k_f+ 3 k_{ISC}+k_0+k_-+k_+\\
C &=(k_{ex}+k_f+ 3 k_{ISC})(k_0+k_-+k_+)\\&\qquad+3 k_{ex} k_{ISC}+ k_0 k_-+k_0 k_++k_- k_+\\
D &= (k_{ex}+k_{f}+3 k_{ISC})(k_0 k_-+k_0 k_++k_- k_+)\\&\qquad+2 k_{ex} k_{ISC}(k_0+k_-+k_+)+k_0\:k_-\:k_+\\
E &=(k_{ex}+k_{f}+3 k_{ISC})(k_0\:k_-\:k_+)\\&\qquad+ 2 k_{ex} k_{ISC}(k_0 k_-+k_0 k_++k_- k_+)
\end{split}
\end{equation}
Additionally, we define the rate of detected photons as,
\begin{equation}\label{eq:rate_photons}
\begin{split}
R &= \frac{k_{f}\:k_{ex}\:k_0\:k_-\:k_+\:\eta}{E}
\end{split}
\end{equation}
where $\eta$ is the collection efficiency of the optical setup. Using \textit{Vieta's} formula \cite{weisstein2002vieta}, we can relate the coefficients of the polynomial to sums and products of its roots.
\begin{equation}\label{eq:poly_roots}
\begin{split}
-B &=\lambda_1+\lambda_2+\lambda_3+\lambda_4\\
C &=(\lambda_1 \lambda_2)+(\lambda_1 \lambda_3)+(\lambda_1 \lambda_4)\\&\qquad+(\lambda_2 \lambda_3)+(\lambda_2 \lambda_4)+(\lambda_3 \lambda_4)\\
-D &=(\lambda_1 \lambda_2 \lambda_3)+(\lambda_1 \lambda_2 \lambda_4)+(\lambda_1 \lambda_3 \lambda_4)\\&\qquad+(\lambda_2 \lambda_3 \lambda_4)\\
E &=\lambda_1\:\lambda_2\:\lambda_3\:\lambda_4
\end{split}
\end{equation}

\subsection{Extraction of $k_i$ parameters}

Using \ref{eq:poly_coeff} and \ref{eq:poly_roots}, the rates $ k_{T_+}$, $k_{T_-} $ and  $k_{T_0} $ can be expressed as a combination of the decay rates and their pre-exponential terms. The extracted triplet depopulation rates are shown in fig.~4~(d). After some substitutions, $k_{ex}$ can be expressed as
\begin{equation}\label{eq:kex_expr}
\begin{split}
k_{ex}^2&- B_n\:k_{ex} + C_n + R_n =0\\
\text{where,}\\
B_n &=B - k_0-k_--k_+ \\
C_n &=C - (k_0 k_-+k_0 k_++k_- k_+)\\&\qquad-B_n(k_0+k_-+k_+) \\
R_n &=\frac{R\:E}{ k_0\:k_-\:k_+ \eta} 
\end{split}
\end{equation}
Thus $k_{ex}$ can be written as,
\begin{equation}\label{eq:kex_soln}
\begin{split}
k_{ex}=\frac{1}{2}(B_n-\sqrt{{B_n}^2 - 4 (C_n+R_n)}
\end{split}
\end{equation}
The extracted values of $k_{ex}$ as a function of the measured excitation power is plotted in fig.~4~(e). The other rates are expressed similarly,
\begin{equation}\label{eq:rates_soln}
\begin{split}
 k_{f}=\frac{R_n}{k_{ex}},\quad
 k_{ISC}=\frac{C_n}{k_{ex}}
\end{split}
\end{equation}
The observed intensity dependence of $k_{ISC}$ is phenomenological included in the rate equation as
\begin{equation}
    \label{eq:rateeq2}
    \begin{split}  
     \dot{\text{S}}_\text{0} &= -k_{ex} \text{S}_\text{0} + k_f \text{S}_\text{1} + k_0 \text{T}_0 + k_- \text{T}_- + k_+ \text{T}_+ \\
     \dot{\text{S}}_\text{1} &= k_{ex} \text{S}_\text{0} - k_f \text{S}_\text{1} - k_{ISC} \text{S}_\text{1} - k_{ex}\:\beta\:\text{S}_\text{1} \\
     \dot{\text{T}}_+ &= \frac{k_{ISC}}{3} \text{S}_\text{1} - k_+ \text{T}_+ \\
     \dot{\text{T}}_- &= \frac{k_{ISC}}{3} \text{S}_\text{1} - k_- \text{T}_- \\
     \dot{\text{T}}_0 &= \frac{k_{ISC}}{3} \text{S}_\text{1} - k_0 \text{T}_0 + k_{ex}\:\beta\:\text{S}_\text{1}
     \end{split}
\end{equation}
For simplicity, we have added the intensity dependent population transfer pathway only to the long lived triplet sub-level. Using the modified rate equations, we obtain $k_{ISC}=\frac{C_n}{k_{ex}}+\beta k_{ex}$, where $\beta$ describes the absorption cross section. This second absorption cross section is estimated from the slope of $k_{ISC}$.

\label{sec:rates-fitting}

\section{Optical Dipole and Spin-Spin Tensor Orientations}
\label{sec:electronic-structure}
Here we consider all possible HOMO and LUMO pairs that can be formed from the vacancy-centered MOs in $C_{2v}$ or $C_{1h}$ symmetry. By evaluating the optical dipole moment of the ground to first excited singlet transition and the spin-spin tensor components of the intermediate triplet, we identify those pairs that are consistent experiment.

\begin{figure}[]
    \centering
    \includegraphics{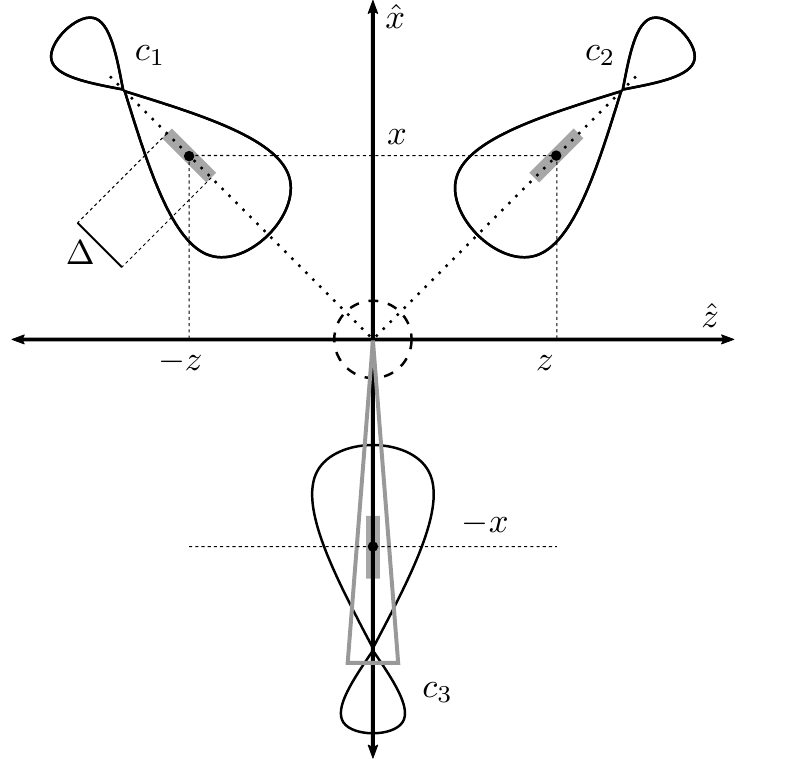}
    \caption{A sketch of the nuclear geometry and atomic orbitals of the viewed along the $\hat y$ direction. The vacancy $V$, and  atomic orbitals, $c_1$ and $c_2$, are in the the $xz$-plane. The $c_3$ orbital lies in the out-of-plane direction. The mean position of the electron density is marked by a black circle and its in-plane positions are labelled by $\expval{x}$ and $\expval{z}$. The variance of $c_1$ with along the orientation of the bond is marked by $\Delta$.}
	\label{fig:geom}
\end{figure}

To begin let the the HOMO be denoted as $a$ and the LUMO as $b$. The ground configuration is $a^2$ and the first excited configuration is $ab$. The corresponding orbital states are: $\phi_g = aa$ and $\phi_e = ab$, respectively. The term states of these configurations are formed by taking direct products of the orbital state with their associated spin states and then applying a Slater determinant to enforce electron interchange anti-symmetry. The resultant electronic states, $\Phi_{S,m_s;\Gamma}$, have well defined total spin $S$, spin projection $m_s$, and orbital symmetry. Since the defect has low symmetry, like $C_{1h}$ and $C_{2v}$, the orbital symmetry is given by the simple product of the symmetries its constituent orbital states. These are defined explicitly as
\begin{align}
    \ket{\Phi_{0,0;\phi_g}} &= \sket{a \bar a}\nonumber\\
    \ket{\Phi_{0,0;\phi_e}} &= \frac{1}{\sqrt{2}}\pqty{\sket{a\bar b} - \sket{\bar ab}} \nonumber\\
    \ket{\Phi_{1,m_s;\phi_e}} &= 
        \begin{cases}
            \sket{ab} & m_s=-1 \\
            \frac{1}{\sqrt{2}}\pqty{\sket{a\bar b} + \sket{\bar ab}} & m_s=0 \\
            \sket{\bar{a}\bar{b}} & m_s=1
        \end{cases}
\end{align}
where $\sket{\cdot}$ denotes a Slater determinant.
In this basis, we can exploit orbital symmetry to simplify evaluation of the transition dipole matrix elements. First, the multi-electron electric-dipole matrix elements can be written in terms of the single MOs as follows
\begin{align}
\matrixel{\Phi_{0,0;\phi_g}}{\hat{d}}{\Phi_{0,0;\phi_e}} &= \frac{1}{\sqrt{2}} \sbra{a\bar a} \hat{d} \pqty{\sket{a\bar b} - \sket{\bar ab}} \nonumber\\
 &= \frac{2}{\sqrt{2}} \matrixel{a}{\hat{d}}{b}
\end{align}
where $\hat d = e\pqty{x \hat x + y \hat y + z \hat z}$ is the electric dipole operator where we have defined the ST1 coordinate system such that $\hat z||[110]$ and $\hat x$ is in the $z$-symmetry plane of the defect. Now it suffices to estimate this matrix element for all unique HOMO and LUMO pairs. In \cref{tab:selection-rules}, we apply the symmetry selection rules for the electric dipole operator to determine the non-zero matrix elements for $C_{2v}$ and $C_{1h}$, respectively. 
\begin{table}[]
\centering
\begingroup
\setlength{\tabcolsep}{10pt} 
\renewcommand{\arraystretch}{1.5} 
\begin{tabular}{c | c  c | c c}
\multicolumn{1}{c|}{} & \multicolumn{2}{c|}{$C_{1h}$} & \multicolumn{2}{c}{$C_{2v}$} \\
$\Gamma$ & $\hat d$ & $\hat V_{ss}$ & $\hat d$ & $\hat V_{ss}$ \\
\hline
$A_{1}$  & $x$, $z$  & $x^2$, $y^2$, $z^2$, $xz$ & $x$ & $x^2$, $y^2$, $z^2$\\
$B_{1}$  & $x$, $z$  & $x^2$, $y^2$, $z^2$, $xz$ & $z$ & $xz$\\
$B_{2}$  &  $y$      & $xy$, $yz$ & $y$ & $xy$ \\ 
$A_{2}$  &  $y$      & $xy$, $yz$ & $-$ & $yz$\\
\end{tabular}
\endgroup
\caption{The linear and quadratic operator symmetry selection rules for $\matrixel{A_1}{\hat O}{\Gamma}$ where $\hat O$ is either the electric-dipole operator $\hat d$ or the orbital spin-spin tensor $\hat D_{ij}$.}
\label{tab:selection-rules} 
\end{table}
In $C_{1h}$, we cannot distinguish $x$ and $z$ orientated dipoles directly from symmetry. Thus, for some HOMO/LUMO pairs we need to directly estimate the matrix elements. We do this by expanding the MOs in terms of their atomic orbitals, neglecting orbital overlap, and applying geometric arguments to simplify the matrix elements. For example, consider the HOMO/LUMO pair $(a_1,b_1)$. The dipole matrix element is
\begin{align}
    \matrixel{a_1}{\hat{d}}{b_1} &= e \matrixel{c_1 + c_2}{x \hat x + z \hat z}{c_1 - c_2} \nonumber\\
     &= e(\bqty{\expval{x}_1-\expval{x}_2}\hat x + \bqty{\expval{z}_1 -\expval{z}_2}\hat z \nonumber\\
     &\;\;\;\; -2\bqty{\matrixel{c_1}{x}{c_2}\hat x  + \matrixel{c_1}{z}{c_2}\hat z}) \nonumber \\
     &\approx e(\bqty{\expval{x}_1-\expval{x}_2}\hat x + \bqty{\expval{z}_1 -\expval{z}_1}\hat z) \nonumber
\end{align}
where $\expval{z}_1$ is the expectation value of $z$ with respect to the $c_1$ atomic orbital. As depicted in \cref{fig:geom}, the $c_1$ and $c_2$ orbitals are both centered in the $xz$-plane at the same mean $x$ position and anti-symmetric mean $z$ positions. Hence, the orbital integrals are related such that: $\matrixel{c_1}{x}{c_1}=\matrixel{c_2}{x}{c_2}$ and $\matrixel{c_1}{z}{c_1}=-\matrixel{c_2}{z}{c_2}$. This means that the the $x$-component of the integral cancels out, leaving
\begin{align}
    \matrixel{a_1}{\hat{d}}{b_1}
    &\approx 2 e\expval{z}_1 \hat z
\end{align}
An analogous argument applied to the other ambiguous HOMO/LUMO pair: $(a_1',b_1)$, yields $\matrixel{a_1'}{\hat{d}}{b_1} \approx 2 e\expval{z}_1 \hat z$.

Next, to calculate the orientation of the major spin axis of the defect, we need to evaluate the components of the triplet level's spin-spin interaction. The zero-field spin Hamiltonian of the triplet is
\begin{align}
\hat H =\mathbf S\cdot\mathbf{D}\cdot \mathbf S
\end{align}
where $\mathbf{S}=S_x \hat{x}+ S_y \hat{y}+S_z \hat{z}$ are the $S=1$ spin operators and $\mathbf{D}$ is the spin-spin tensor with components $D_{ij}=\sbra{\phi_e}\hat{D}_{ij}\sket{\phi_e}$ and
\begin{align}
    \hat{D}_{ij} = 
C \bqty{\frac{\delta_{ij}}{r^3} - \frac{3 r_{i} r_{j}}{r^5}}
\end{align}
where $C=3\mu_0g_e^2\mu_B^2/16\pi h$, $\mu_0$ is the vacuum permeability, $g_e\approx2$ is the free electron g-factor and $\mu_B$ is the Bohr magneton. The vector distance between the two electrons is $\mathbf r = \mathbf r_2-\mathbf r_1$, the magnitude of this vector is defined as $r$, and $r_i$ is its components in the $x$, $y$, or $z$ directions. We apply symmetry selection rules to simplify the evaluation of these tensor components in \cref{tab:selection-rules}. In $C_{2v}$, the only quadratic operators that transform as the trivial representation are $x^2$, $y^2$, and $z^2$. These, in addition to $xz$, form the basis of quadratic operators in $C_{1h}$ that transform as the trivial representation. Hence, it suffices to calculate  the two electron integrals $\expval{D_{xx}}$, $\expval{D_{yy}}$, $\expval{D_{zz}}$ and $\expval{D_{xz}}$ for each HOMO/LUMO pair. 

As we did with the dipole moment, we expand the non-zero terms in the basis of atomic orbitals. The resulting expression is still too difficult to integrate directly. The dominant contributions to the tensor component can be evaluated by extending the semi-classical approximation introduced in \citet{Doherty_NV_Under_Pressure}. The approximation corresponds to the replacement
\begin{align}
    \expval{\frac{\delta_{ij}}{r^3} - \frac{3 r_i r_j}{r^5}}_{\alpha\beta} \rightarrow 
    \frac{1}{\abs{\expval{\mathbf r_1}-\expval{\mathbf r_2}}^3} - \frac{3\pqty{\expval{r_i}\expval{r_j} - \Delta_{ij}}}{\abs{\expval{\mathbf r_1}-\expval{\mathbf r_2}}^5}
\end{align}
where $\alpha, \beta$ are atomic orbitals, such that $\expval{\mathbf r_1}=\matrixel{\alpha}{\mathbf r_1}{\alpha}$ and $\expval{\mathbf r_2}=\matrixel{\beta}{\mathbf r_2}{\beta}$ is the mean position of the electrons occupying the $\alpha$ and $\beta$ sp$^3$ orbitals. The first term can be interpreted physically as the interaction between electrons at their mean positions. The $\expval{r_i}\expval{r_j}$ component of the second term accounts for the mean relative positions of the electrons. This second term also includes 
\begin{align}
\Delta_{ij}&=\matrixel{\alpha \beta}{r_i r_j}{\alpha \beta}-\matrixel{\alpha \beta}{r_i}{\alpha \beta}\matrixel{\alpha \beta}{r_j}{\alpha \beta} 
\end{align}
which is the co-variance in the electron position along the $ij$ axis. 
These approximate terms can be estimated by applying geometric arguments on a case by case basis for each of the HOMO/LUMO pairs. We can use the same arguments as the electric dipole case to simplify the mean positions. We estimate $\Delta_{ij}$ by considering the electron distribution of each sp$^3$ bond. 
Each atomic orbital has a major axis along the line between the vacancy and the orbital's center. We define two minor axes perpendicular to the major axis, one in-plane and the other out of plane. With these definitions, we note that the co-variance in the major axis of each orbital is much larger than its co-variance in the minor axes. Applying this to our example, the $(a_1,b_1)$ pair, the variance $\Delta_{yy}\approx 0$ and $\Delta_{xz}\approx\Delta_{xx}\approx\Delta_{zz}$. Accordingly, the spin-spin tensor takes the form
\begin{align}
\expval{D}\approx C\begin{pmatrix}
A-B & 0 & B \\
0 & A & 0 \\
B & 0 & -2A+B 
\end{pmatrix} 
\end{align}
where $A=4/\expval{r_z}^3$ and $B=12 \Delta_{xz}/\expval{r_z}^5$. The off-diagonal terms imply that the major spin axis is not exactly in the [110] direction. Rotating the spin-spin tensor around the $\hat{y}$ gives us the offset of spin-quantization major axis, i.e. finding a rotation such that ${M}\pqty{\theta }^{-1}{D}{M}\pqty{\theta}$ is diagonal. Solving this equation and Taylor expanding the result we find that (to first order) $\theta \approx B/\pqty{2 B-3 A}$. Given that the variance is much smaller that the mean, $B<<A$, we claim that $\theta\approx 0$. With this result, we can conclude that the major spin axis and electric dipole of the $(a_1,b_1)$ pair are co-aligned in the $[110]$ ($\hat z$) direction, and the spin minor axis is in the $\hat{x}$ direction. 
\begin{table}[h]
\centering
\begingroup
\setlength{\tabcolsep}{7pt} 
\renewcommand{\arraystretch}{1.5} 
\begin{tabular}{c | c   c  c }
(HOMO, LUMO)               & $\Gamma$ & Dipole Moment & Spin Axis \\
\hline
 $(a_1 a'_1)$              & $A_{1}$  & $x$       & $x$  \\
 $(a_1 b_1)$               & $B_{1}$  & $z$       & $z$ \\
 $(a'_1 b_1)$              & $B_{1}$  & $z$       & $x$ \\
 $(a_1 b_2)$               & $B_{2}$  & $y$       & $x$ \\ 
 $(a'_1 b_2)$              & $B_{2}$  & $y$       & $y$ \\ 
 $(b_1 b_2)$               & $A_{2}$  & $y^\dagger$        & $x$ \\
\end{tabular}

$^\dagger$This dipole moment is not allowed in $C_{2v}$ in symmetry.
\endgroup
\caption{A table of the properties of each HOMO/LUMO pair. It includes the symmetry of the excited electronic state, the orientation of the dipole moment and the major spin quantization axis for each pair. 
}
\label{tab:orientations} 
\end{table}

The orientations electric-dipole moment and spin-quantization axis for each of the HOMO/LUMO pairs in \cref{tab:orientations}. In the defect coordinate system $\hat z$ and $\hat y$ are the in-equivalent $[110]$ directions. Hence, the only HOMO/LUMO pair which are consistent with experiment are: $(a_1,b_1)$ and $(a_1',b_2)$ because the electric-dipole moment and spin-quantization axis are in the $\hat z$ $\hat y$ directions. Further, our estimates predict that in both configurations the electric-dipole moment and spin quantization are co-linear to first order.

\bibliographystyle{unsrtnat}
\bibliography{refs}